\documentclass[aps,prl,twocolumn]{revtex4}
\usepackage{amsmath}
\usepackage{graphicx}
\begin{document}

\title{Phenomenology of Current-Induced Dynamics in Antiferromagnets}

\author{Kjetil M. D. Hals$^1$, Yaroslav Tserkovnyak$^2$, and Arne Brataas$^1$}
\affiliation{$^1$ Department of Physics, Norwegian University of Science and Technology, NO-7491, Trondheim, Norway \\
$^2$Department of Physics and Astronomy, University of California, Los Angeles, California 90095, USA}

%%%%%%%%%%%%%%%%%%%%%%%%%%%
\begin{abstract}
We derive a phenomenological theory of current-induced staggered magnetization dynamics in antiferromagnets. The theory captures the reactive and dissipative current-induced torques and the conventional effects of magnetic fields and damping.  A Walker ansatz describes the dc current-induced domain-wall motion when there is no dissipation. If magnetic damping and dissipative torques are included, the Walker ansatz remains robust when the domain-wall moves slowly. As in ferromagnets, the domain-wall velocity is proportional to the ratio between the dissipative-torque and the magnetization damping. In addition, a current-driven antiferromagnetic domain-wall acquires a net magnetic moment.
\end{abstract}
%%%%%%%%%%%%%%%%%%%%%%%%%%%
\maketitle

%%% New commands %%%
\newcommand{\eq}  {  \! = \!  }
\newcommand{\keq} {\!\! = \!\!}
\newcommand{\kadd}{  \! + \!  }
\newcommand{\ksim}{\! \sim \!}

%%%%%%%%%%%%%%%%%%%%%%%%%%%
%\section{Introduction}
%%%%%%%%%%%%%%%%%%%%%%%%%%%
In ferromagnets, a spin-polarized current can be used to manipulate magnetization via the exchange interaction. A misalignment between the polarization of the current and the local magnetization direction causes a spin-transfer torque (STT) on the magnetization because of noncollinear spins that precess within the ferromagnet. This effect was first theoretically predicted by Slonczewski and Berger~\cite{Slonczewski_Berger} and has since garnered abundant experimental evidence (for a review, see Ref.~\cite{Ralph:jmmm08}). The STT effect is the reciprocal process of the charge currents that are induced by a time-varying magnetic texture~\cite{Tserkovnyak:PRB08}. A promising commercial application of the STT effect utilizes the spin-polarized currents to switch the ferromagnetic layers in spintronic devices, such as in magnetic random-access memory or to induce magnetic precession for use in high-frequency oscillators in wireless communication devices. A limitation of these applications of STT is the high levels of critical currents that are required to switch the direction of the magnetization. 

Antiferromagnets are ordered spin systems in which the magnetic moments of all electrons in each unit cell compensate for each other in equilibrium. In spintronic devices, antiferromagnets are commonly used to pin ferromagnetic layers via the exchange bias effect. Recent theoretical~\cite{AF_theory,Swaving:cm09} and experimental~\cite{Urazhdin:PRL07} works indicate, however, that current-induced torque effects also appear in antiferromagnets. Antiferromagnets also share another transport property with ferromagnets; namely, the anisotropic magnetoresistance (AMR) effect~\cite{AF_AMR}. The antiferromagnetic AMR effect allows for a detailed experimental study of the current-induced switching of the antiferromagnetic layers and the motion of the spatially dependent antiferromagnetic textures. Therefore, antiferromagnets can replace ferromagnets for use in spintronics devices. The applicability of antiferromagnets depends on the critical currents that are needed to manipulate the staggered magnetization as well as on the resistance variation that such a reorientation of the magnetization induces. 

Magnetization dynamics in ferromagnets is described by the Landau-Lifshitz-Gilbert (LLG) equation, which has been extended to include STT~\cite{Tserkovnyak:jmmm08}. In magnetic textures, a spin-polarized current contributes two different terms to the LLG equation of motion: a reactive torque term and a dissipative torque term~\cite{Tserkovnyak:jmmm08}. The reactive torque term preserves the macroscopic time-reversal invariance of the equation. This term arises from the out-of-equilibrium spin density that is induced by drifting electrons, which have spins that are adiabatically guided by the magnetic texture. The dissipative torque breaks the time-reversal symmetry and arises from the spin-dephasing processes. Except for those systems with strong spin-orbit coupling~\cite{Hals:prl09}, the magnitude of the dissipative torque is typically much smaller than the reactive torque, but it is not less important~\cite{Tserkovnyak:jmmm08}. In some ferromagnetic domain-wall systems, the dissipative torque dictates the current-induced domain-wall velocity~\cite{Tserkovnyak:jmmm08}.

The effect of the reactive STT on the staggered magnetization in antiferromagnets was recently discussed in Ref.~\cite{Swaving:cm09}. The torque has the same physical origin as that in ferromagnets. Based on the form of the dissipative STT in ferromagnets, Ref.~\cite{Swaving:cm09} also makes an educated guess about the form of the dissipative STT in antiferromagnets.  

In the present paper, we develop a general phenomenology that describes the coupled dynamics of currents and the staggered order parameter in isotropic antiferromagnets to the lowest order in spin-texture gradients and precession frequency. The antiferromagnet is treated within the exchange approximation. For the lowest order in relativistic interactions, the exchange forces only depend on the relative orientation of the spins. This approximation is a good starting point for many conventional ferromagnets and antiferromagnets~\cite{Andreev:SPU80}, including disordered systems. In these systems, impurities couple to the spin degrees of freedom through random magnetic moments or spin-orbit coupling, but impurity averaging restores the spin-rotational and sublattice symmetries. We include the effects of damping, external magnetic fields, and reactive and dissipative torque effects.  Our results differ from the postulated form of the dissipative torque in Ref.~\cite{Swaving:cm09}, and we explain why.  We apply our theory to an antiferromagnetic domain-wall system, and find an analytic solution in the low current-density regime. Similar to ferromagnets, we find that the domain-wall velocity is proportional to the ratio between the dissipative-torque and a bulk damping coefficient. An interesting consequence of the current-induced motion is that the domain-wall develops a net magnetic moment. Current-induced staggered magnetization dynamics can thus be observed in two ways:  via the AMR effect and via the out-of-equilibrium net magnetic moment.

%%%%%%%%%%%%%%%%%%%%%%%%%%%
%\section{Theory}
%%%%%%%%%%%%%%%%%%%%%%%%%%%

Our phenomenology is based on the theory of insulating antiferromagnets~\cite{LL_StatPhys2}, which is extended to take into account the current flow. An important aspect of our phenomenology is the exchange approximation that implies that the total energy is invariant during the simultaneous rotation of all the magnetic moments~\cite{LL_StatPhys2}. Subsequently, when considering the current-induced domain-wall motion, we include the magnetic anisotropy phenomenologically in the free energy, considering that these anisotropy energies are very small, e.g., on the scale of the critical temperature. 

For clarity, we restrict our treatment to systems in which each unit cell in the crystal lattice contains two equivalent magnetic sites. In this situation, the antiferromagnet consists of two sublattices with magnetic moment densities $\mathbf{m}_1(\mathbf{r},t)$ and $\mathbf{m}_2(\mathbf{r},t)$, such that the total magnetization is $\mathbf{m}(\mathbf{r},t) =  \mathbf{m}_1(\mathbf{r},t) + \mathbf{m}_2(\mathbf{r},t)$ and the antiferromagnetic order parameter is $\mathbf{l}(\mathbf{r},t) =  \mathbf{m}_1(\mathbf{r},t) - \mathbf{m}_2(\mathbf{r},t)$. In equilibrium and in the absence of magnetic fields and textures, $\mathbf{m}$ vanishes, and $\mathbf{l}$ is finite and homogenous. In the following, we allow the antiferromagnet to become distorted into metastable textured states, such as domain-walls or vortices, but we require that the texture is smooth on the scale of relevant microscopic length scales. The texture is parameterized by a slowly varying unit vector $\mathbf{n}(\mathbf{r},t)\equiv\mathbf{l}(\mathbf{r},t)/ l$ ($l\equiv | \mathbf{l}(\mathbf{r},t) |$). Assuming stiff antiferromagnetic ordering, the longitudinal dynamics of $\mathbf{l}$ can be neglected so that the slow dissipative dynamics of the system are fully described by the directional N{\'e}el field $\mathbf{n}(\mathbf{r},t)$ along with the transverse magnetization $\mathbf{m}(\mathbf{r},t)$, which physically corresponds to the small relative canting of the magnetic sublattices. Constructing the phenomenological equations of motion well below the N{\'e}el temperature, we thus impose the constraints $|\mathbf{n}|=1$ and $\mathbf{m}\cdot\mathbf{n}=0$. This starting point is analogous to Haldane's mapping of the long-wavelength antiferromagnetic action into the nonlinear sigma model \cite{Auerbach}.

In addition to rotational invariance, the exchange approximation requires that the free energy and the equations of motion are invariant  under the exchange of the two sublattices~\cite{LL_StatPhys2}, i.e., that they are invariant under the   transformations 
$\mathbf{n}(\mathbf{r},t)\mapsto -\mathbf{n}(\mathbf{r},t)$ and $\mathbf{m}(\mathbf{r},t)\mapsto \mathbf{m}(\mathbf{r},t)$. The leading-order free energy that satisfies the appropriate symmetry requirements is thus~\cite{LL_StatPhys2}
\begin{equation} 
F=\int d\mathbf{r}\left[ \frac{1}{2}a\mathbf{m}^2   + \frac{A}{2}\sum_{i= {x,y,z}}\left( \partial_i \mathbf{n} \right) ^2   
- \mathbf{H}\cdot\mathbf{m} \right]\,.
\label{FreeEnergy} 
\end{equation}
Here, we expanded the free energy to the second order in the gradients and the magnetization field $\mathbf{m}(\mathbf{r},t)$, which is coupled to an external magnetic field $\mathbf{H}$.
The equations of motion for $\mathbf{m}(\mathbf{r},t)$ and $\mathbf{n}(\mathbf{r},t)$ are found by expanding their slow dynamics to the lowest order in the effective fields $\mathbf{f}_n\equiv-\delta_{\mathbf{n}}F=A\mathbf{n}\times(\nabla^2\mathbf{n}\times\mathbf{n})-\mathbf{m}(\mathbf{H}\cdot\mathbf{n})$ and $\mathbf{f}_m\equiv-\delta_{\mathbf{m}}F=-a\mathbf{m}+\mathbf{n}\times \left(  \mathbf{H}\times\mathbf{n} \right)$. To enforce the constraints $|\mathbf{n}|=1$ and $\mathbf{m}\cdot\mathbf{n}=0$, we calculated the variational derivatives $\delta_{\mathbf{m}}F$ by varying $\mathbf{m}$ normal to a fixed $\mathbf{n}$ and $\delta_{\mathbf{n}}F$ by parallel transporting $\mathbf{m}$ on the sphere that is parameterized by $\mathbf{n}$. In the absence of electric currents, we obtain~\cite{foot} 
\begin{align} 
\dot{\mathbf{n}} &=(\gamma\mathbf{f}_m- G_1 \dot{\mathbf{m}})\times\mathbf{n}\,, \label{LLG1} \\
\dot{\mathbf{m}} &=(\gamma\mathbf{f}_n-G_2\dot{\mathbf{n}})\times\mathbf{n}+(\gamma\mathbf{f}_m- G_1\dot{\mathbf{m}})\times\mathbf{m}\,,
\label{LLG2} 
\end{align}
where $\gamma$ is the effective gyromagnetic ratio. The dissipation power $P\equiv\dot{\mathbf{n}}\cdot\mathbf{f}_n+\dot{\mathbf{m}}\cdot\mathbf{f}_m=(G_1/\gamma)\dot{\mathbf{m}}^2+(G_2/\gamma)\dot{\mathbf{n}}^2\geq0$ requires that $G_{1,2}/\gamma\geq0$. The nondissipative equations with $G_{1,2}=0$ are derived in the linearized regime in Ref.~\cite{LL_StatPhys2}. In addition to the appending dissipation, we have also added the second term in Eq.~(\ref{LLG2}), which is quadratic in small deviations from the equilibrium, to enforce the constraint $\mathbf{m}\cdot\mathbf{n}=0$. Note that such a term naturally appears if one constructs the antiferromagnetic equations of motion out of the ferromagnetic LLG equations of the constituent magnetic sublattices. Eq.~\eqref{LLG1}-\eqref{LLG2} can be reduced to a single equation for the N{\'e}el field (without dissipation): $\mathbf{n}\times \ddot{\mathbf{n}}=\gamma^2 a \mathbf{n}\times \left[ A\nabla^2\mathbf{n} - \mathbf{H}\left( \mathbf{H}\cdot\mathbf{n} \right)/a + \dot{\mathbf{H}}\times\mathbf{n}/ \gamma a  \right] - 2\gamma \left( \mathbf{H}\cdot\mathbf{n} \right)\dot{\mathbf{n}} $. This equation agrees with the equation that is derived in Ref.~\cite{Andreev:SPU80} from the Lagrangian density $\mathcal{L} = \left( \dot{\mathbf{n}}/\gamma -\mathbf{H\times n}  \right)^2/2a - A\left( \nabla\mathbf{n} \right)^2/2$. 

We also make use of the linearized equations in the Landau-Lifshitz form:
\begin{equation}
\dot{\mathbf{n}}= \tilde{\gamma}(\mathbf{f}_m\times \mathbf{n} + G_1\mathbf{f}_n)\,,\,\,\,\dot{\mathbf{m}}= \tilde{\gamma}(\mathbf{f}_n\times \mathbf{n} + G_2 \mathbf{f}_m)\,, \label{LL} 
\end{equation}
where $\tilde{\gamma} \equiv \gamma / \left( 1+G_1 G_2 \right)$. The Onsager reciprocity relations between the two fields require the gyromagnetic ratios to be the same in the two equations (see below).

The equations~\eqref{LLG1} and \eqref{LLG2} describe the evolution of an electrically open antiferromagnet. Next, we include the effect of itinerant electrons on the long-wavelength dynamics of $\rm\mathbf{n}(\mathbf{r},t)$ and $\rm\mathbf{m}(\mathbf{r},t)$ by adding torque terms that arise from the currents that are induced by an external electric field $\mathbf{E}$. To this end, we are guided by the rotational symmetry requirements and the Onsager reciprocity relations. The Onsager reciprocity relations apply to a system that is described by several parameters $\{ q_i|i=1,\ldots,N\}$ for which the rate of change $\dot{q}_i$ is induced by the thermodynamic forces $f_i\equiv-\partial_{q_i}F$, and state that the off-diagonal linear response coefficients in the equations $\dot{q}_i=\sum_{j=1}^{N} L_{ij}f_j$ are related by $L_{ij}(\mathbf{H},\mathbf{M})= \epsilon_i\epsilon_jL_{ji}(-\mathbf{H},-\mathbf{M})$, where $\epsilon_i=1$ ($\epsilon_i=-1$) if $q_i$ is even (odd) under time reversal. Here, $\mathbf{M}$ represents any possible equilibrium magnetic order. The fields that describe the collective magnetic dynamics in the antiferromagnet are $\mathbf{n}(\mathbf{r},t)$ and $\mathbf{m}(\mathbf{r},t)$, and the associated conjugate forces are $\mathbf{f}_n$ and $\mathbf{f}_m$, respectively. In the diffusive regime, the charge transferred by the current-density $\boldsymbol{\jmath}$ is conjugate to the electric field such that $\mathbf{f}_q= \mathbf{E}$. The response coefficients that are needed are the response matrices $L_{\mathbf{n},\mathbf{q}}$ and $L_{\mathbf{m},\mathbf{q}}$, which describe the dynamics of  $\mathbf{n}$ and $\mathbf{m}$ that are induced by the electric field. Because the magnetization is odd and the charge is even under the time reversal, Onsager's theorem implies that 
$L_{n_i (m_i),q_j}( \mathbf{n}, \mathbf{m})= -L_{q_j,n_i (m_i)}( -\mathbf{n}, -\mathbf{m})$,  where $L_{\mathbf{q},\mathbf{n} (\mathbf{m})}$ describes the charge currents that are pumped by $\mathbf{f}_n$ ($\mathbf{f}_m$), and the $(\pm\mathbf{n},\pm\mathbf{m})$ arguments denote an equilibrium texture.

To derive the STT terms, it is convenient to begin by phenomenologically constructing the magnetically pumped charge current $\boldsymbol{\jmath}^{\rm pump}$, which yields $L_{\mathbf{q},\mathbf{n} (\mathbf{m})}$, and then invoke Onsager's theorem to obtain  $L_{\mathbf{n}(\mathbf{m}),\mathbf{q}}$. For the lowest order of the space-time gradients and the magnetization field $\mathbf{m}$, we can write three pumping terms that satisfy the appropriate exchange and spatial symmetries: 
$\mathbf{n}\cdot \left( \dot{\mathbf{m}}\times\partial_i\mathbf{n} \right) $, $\dot{\mathbf{n}}\cdot\partial_i\mathbf{n}$, and $\mathbf{n}\cdot \left( \dot{\mathbf{n}}\times\partial_i\mathbf{m} \right) $. However, because the last term is quadratic in the small deviations from an equilibrium state (in the absence of magnetic fields), we disregard it in the following. Thus, the leading-order phenomenological expression for the pumped charge current is as follows:
\begin{align} 
\jmath^{\rm pump}_i/\sigma &= \eta\,\mathbf{n}\cdot\left( \dot{\mathbf{m}}\times\partial_i\mathbf{n}\right)  +  \beta\, \dot{\mathbf{n}}\cdot \partial_i\mathbf{n}  \nonumber \\
& \hspace{-1cm}= \tilde{\gamma}\left[( \eta + G_1 \beta ) \partial_i \mathbf{n}\cdot \mathbf{f}_n +( \beta - G_2\eta )\mathbf{n}\times\partial_i \mathbf{n} \cdot \mathbf{f}_m\right]\,,
\label{PumpedCurrent} 
\end{align}
where we have utilized Eq.~\eqref{LL} and have scaled the current density with the conductivity $\sigma$.
Here, $\eta$ ($\beta$) is a phenomenological parameter. Later, it becomes clear that $\eta$ ($\beta$) parameterizes the adiabatic (non-adiabatic) torque because the term is even (odd) under time reversal. 
Eq.~\eqref{PumpedCurrent} yields the response coefficients
$L_{q_i,n_j}=\sigma \tilde{\gamma}(\eta + G_1 \beta) \partial_i n_j$ and $L_{q_i,m_j}= \sigma\tilde{\gamma}( \beta - G_2\eta )(  \mathbf{n}\times\partial_i \mathbf{n})_j$. 
Using the Onsager reciprocity relations and Ohm's law for the drift current ($\boldsymbol{\jmath}=\sigma\mathbf{E}$), leads to the  
STT terms $\boldsymbol{\tau}_n= \tilde{\gamma}( \eta+ G_1 \beta ) ( \boldsymbol{\jmath}\cdot\boldsymbol{\nabla}) \mathbf{n} $ and  
$\boldsymbol{\tau}_m= -\tilde{\gamma}( \beta - G_2\eta) \mathbf{n}\times( \boldsymbol{\jmath}\cdot\boldsymbol{\nabla}) \mathbf{n}$ for the N\'eel and the magnetization field, respectively, which are added on the right side of the equations of motion in Eq.~\eqref{LL}. Transforming these torques back to the LLG form of the equations,  i.e., Eqs.~(\ref{LLG1}) and (\ref{LLG2}), yields the following:    
\begin{align}
\dot{\mathbf{n}} &=  (\gamma\mathbf{f}_m-G_1\dot{\mathbf{m}})\times\mathbf{n} + \eta\gamma( \boldsymbol{\jmath}\cdot\boldsymbol{\nabla}) \mathbf{n} \,, \label{HTB1} \\
\dot{\mathbf{m}} &= \left[\gamma \mathbf{f}_n - G_2\dot{\mathbf{n}} + \beta\gamma\left( \boldsymbol{\jmath}\cdot\boldsymbol{\nabla}\right) \mathbf{n}\right]\times\mathbf{n}+\boldsymbol{\tau}_{nl} \,.
\label{HTB2} 
\end{align}
$\boldsymbol{\tau}_{nl}=(\gamma\mathbf{f}_m - G_1\dot{\mathbf{m}})\times\mathbf{m}-\eta\gamma\left[\mathbf{m}\cdot(\boldsymbol{\jmath}\cdot\boldsymbol{\nabla})\mathbf{n}\right]\mathbf{n}$ 
are the simplest nonlinear terms that are added here to enforce the constraint $\mathbf{m}\cdot\mathbf{n}=0$. We disregard such higher-order terms in the following.

From now on we use the simplified notation for the effective forces~\cite{foot}, which allows to more readily solve for $\mathbf{m}$ in terms of $\mathbf{n}$.
Combining Eqs.~(\ref{HTB1}) and (\ref{HTB2}), we see that the magnetization field is fully determined by the order parameter $\mathbf{n}$ and its dynamics:
\begin{equation}
\mathbf{m} =\frac{1}{a}\left[\mathbf{n} \times \mathbf{H} + \frac{1}{\tilde{\gamma}}\dot{\mathbf{n}} - G_1\mathbf{f}_n
- (\eta+G_1\beta)( \boldsymbol{\jmath}\cdot\boldsymbol{\nabla}) \mathbf{n}\right]\times\mathbf{n}\,.
\label{mExpression} 
\end{equation}
Substituting this into Eq.~(\ref{HTB2}) allows us to derive a closed equation for the N{\'e}el field to the linear order in the out-of-equilibrium deviations $\mathbf{m}$, $\partial_t\mathbf{n}$, $\boldsymbol{\jmath}$, and $\mathbf{H}$~\cite{foot2}:
\begin{align}
\ddot{\mathbf{n}}/\tilde{\gamma}=&- \mathbf{n}\times\dot{\mathbf{H}} + G_1\dot{\mathbf{f}}_n + (\eta + G_1\beta)(\dot{\boldsymbol{\jmath}}\cdot\boldsymbol{\nabla}) \mathbf{n}\nonumber\\
&+a\left[\gamma\mathbf{f}_n -G_2\dot{\mathbf{n}}+\gamma\beta (\boldsymbol{\jmath}\cdot\boldsymbol{\nabla}) \mathbf{n}\right] \,.
\label{nlsm}
\end{align}
Eq.~\eqref{PumpedCurrent}-\eqref{nlsm} are our main results, which describe a general phenomenological theory of weakly excited current-induced dynamics in conducting antiferromagnets and charge pumping that arises from moving textures. The reactive torque in Eq.~\eqref{nlsm}, which is proportional to $(\eta +G_1\beta )$, was first found in Ref.~\cite{Swaving:cm09}. The dissipative torque term, which is proportional to $a\beta$, and the effects of magnetization damping are new terms that have not been derived before. The consideration of the charge pumping that occurs when moving antiferromagnetic textures is also new. Ref.~\cite{Swaving:cm09} suggests a dissipative torque of the form $\beta \mathbf{n}\times ( \dot{\boldsymbol{\jmath}}\cdot\boldsymbol{\nabla} ) \mathbf{n} $ in Eq. \eqref{nlsm}.  This term breaks the $\mathbf{n}\mapsto -\mathbf{n}$ invariance of the equation and therefore cannot appear in the exchange approximation of equivalent magnetic sublattices.

%%%%%%%%%%%%%%%%%%%%%%%%%%%
%\section{Results and discussion}
%%%%%%%%%%%%%%%%%%%%%%%%%%%
As an application of our theory, we consider an antiferromagnetic domain-wall system and study the current-induced domain-wall motion. For clarity, we set the external magnetic field to equal zero from this point of the paper on. 
Domain-walls can be created in systems with anisotropy, which is added phenomenologically to the free energy as follows:
$F\left[ \mathbf{m}, \mathbf{n}  \right] \rightarrow F\left[ \mathbf{m}, \mathbf{n}  \right] + W\left[ \mathbf{n}  \right]$, where
$W \left[ \mathbf{n}  \right] = \int d \mathbf{r} \left( K_{\bot} n_y^2 /2 - K_{z} n_z^2 / 2 \right) $ is the anisotropy energy ($K_{\bot},K_{z} > 0 $).
A local minima of the above energy functional is a N\'eel wall that rotates in the $xz$ plane, where the local magnetization direction is $n_x = 1/\cosh (z/\lambda_w)$ and $n_z = \tanh (z/\lambda_w)$. $\lambda_w= \sqrt{A/K_z}$ is the domain-wall width. This equilibrium domain-wall texture is denoted by $\mathbf{n}_{0}(\mathbf{r},t)$ below. 

In the following, we study how the domain-wall moves in response to a current along the $z$ axis. Let us first consider the case of no magnetization damping  and $\beta=0$. In this case, it follows from Eq.~\eqref{HTB1}-\eqref{HTB2} that $\mathbf{m}(\mathbf{r},t)=0$ and $\mathbf{n}(\mathbf{r},t)= \mathbf{n}_{0}(z-r_w(t) )$ is an exact solution of the equations, with the domain-wall velocity $\dot{r}_w = -\gamma\eta\jmath $. By including the magnetization damping and the dissipative torque, a local magnetic moment density develops. The torque  
$\boldsymbol{\tau_m}= -\tilde{\gamma}\jmath \left( \beta - G_2\eta  \right) \mathbf{n}\times\partial_z \mathbf{n} $ induces a magnetic moment density along the $y$-axis that should eventually approach a finite value due to the opposite acting damping term $G_2 \mathbf{f}_m $ in Eq.~\eqref{LL}. Thus, to find a stationary solution for Eq.~\eqref{HTB1}-\eqref{HTB2}, we use the ansatz $ \mathbf{n}(\mathbf{r},t)= \mathbf{n}_{0}(z-r_w (t) )$ and 
$\mathbf{m} ( \mathbf{r},t ) = m_0 (t) \mathbf{n}\times \partial_z \mathbf{n}$. Here,  $m_0 (t)$ parameterizes the magnitude of the local magnetic moment density. Substituting these two expressions into Eq.~\eqref{HTB1}-\eqref{HTB2}, produces the following equations for the two parameters $ r_w (t)$ and $m_0(t)$:  
\begin{eqnarray} 
 \dot{r}_w &=& \tilde{\gamma}\left[ a m_0 - \left( \eta + G_1\beta \right)\jmath   \right] ,  
\label{rdot} \\ 
\frac{\dot{m}_0}{\tilde{\gamma}} &=&  \left( G_2\eta  - \beta \right) \jmath   -  G_2 a \left( 1 + \frac{\dot{r}_w n_z }{\tilde{\gamma} G_2 a \lambda_w} \right)  m_0 . \label{mdot}
\end{eqnarray}       
The second term inside the last parenthesis in Eq.~\eqref{mdot} is position-dependent because of $n_z$. When this term is negligible, the ansatz becomes a good approximation of Eq.~\eqref{rdot}-\eqref{mdot}.
This low current-density regime corresponds to systems for which the characteristic intrinsic relaxation time $(\tilde{\gamma} G_2 a )^{-1}$
of the antiferromagnetic system is much smaller than the timescale $\lambda_w/\dot{r}_w$; i.e., the domain-wall moves a small distance as compared to the domain-wall width during the relaxation time. 
In this regime, $m_0$ approaches a finite stable value of $m_0 =  -\left( \beta - G_2\eta \right)\jmath / \left(  G_2 a \right) $, and the domain-wall moves at a constant velocity $\dot{r}_w =  -\gamma \beta \jmath / G_2$.  
Similarly to that for ferromagnets, the velocity is proportional to the ratio between the dissipative torque and a Gilbert damping coefficient. In contrast to ferromagnets, this result is independent of the uniaxial anisotropy $K_{\bot}$ such that the Walker ansatz has a much wider range of applicability. An interesting consequence of the current-induced domain-wall motion is that the system develops a finite magnetic moment in the domain-wall region. This effect may be an alternative to the AMR effect for the measurement of domain-wall motion.

\begin{figure}[ht] 
\centering 
\includegraphics[scale=1.0]{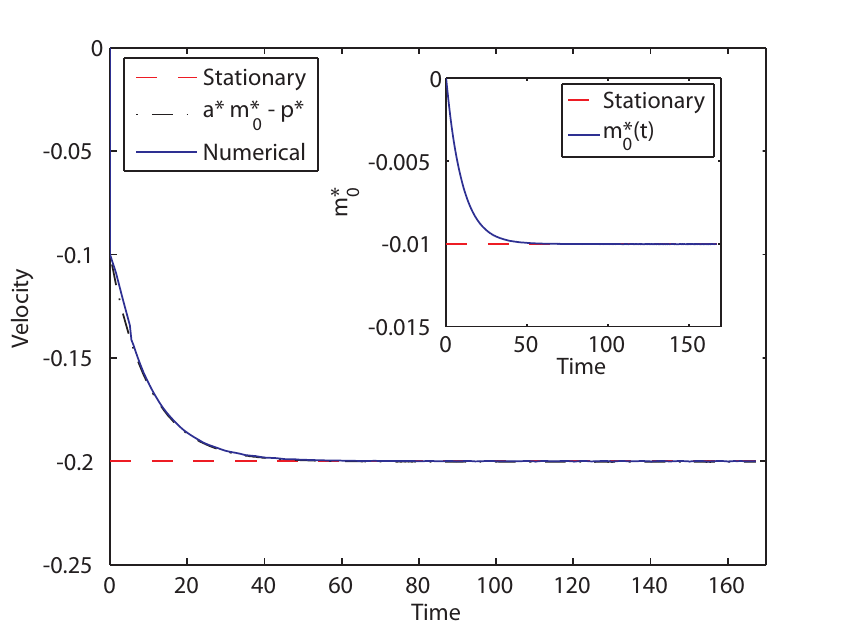} 
\caption{The blue line shows the domain-wall velocity, which was determined using a micromagnetic simulation, as a function of time when a current is applied at $t=0$. The velocity follows the analytic expression in Eq.~\eqref{rdot} and approaches the stationary value $\dot{r}_w =  -\gamma\beta \jmath / G_2$.  Inset: The time evolution of $\rm{m_0}$. The parameter approaches the stationary value $m_0 =  -\left( \beta - G_2\eta \right)\jmath / \left(   G_2 a  \right)$. All of the results are given in dimensionless quantities. $a^{\ast}=a l / A^{\ast}$, and $p^{\ast} = \jmath \left(  \eta + G_1\beta  \right) / \left(  a_l  A^{\ast} \right)$.}
\label{fig:Fig1} 
\end{figure} 

To verify that the system approaches the above stationary solution in the relevant regime,
we conducted a micromagnetic simulation of a one-dimensional system based on Eq.~\eqref{HTB1}-\eqref{HTB2}.
For the numerical calculation, we wrote the equations in a dimensionless form by scaling the $z$ axis with the lattice constant $a_l$ and the time axis with $ \left(\tilde{\gamma}A^{\ast} \right)^{-1}$. Here, $A^{\ast} = A/\left( l a_l^2 \right)$. We considered a domain-wall system with a domain-wall width of $ \lambda_w = 20 a_l$. The anisotropy and damping parameters are 
$a l / A^{\ast}=10$, $K_z /\left( l A^{\ast} \right) = (20)^{-2}$, $K_{\bot}/ \left( l A^{\ast} \right) = 0.1$ and 
$G_1 l = G_2/  l =0.01$. For the STT torque parameters we used 
$\jmath \left(  \eta + G_1\beta  \right) / \left(  a_l  A^{\ast} \right) = 0.1$ and 
$\jmath \left(  \beta - G_2 \eta  \right) / \left(  a_l  A^{\ast} l \right) = 0.001$.
With these values, the stationary solution implies that $m_0^{\ast}= m_0/\left(a_l  l \right) $ will approach $0.01$ and that the wall will move a distance of $2 a_l$ during the relaxation time $(\tilde{\gamma} G_2 a  )^{-1}$. We therefore expect the system to be in the relevant regime for which the stationary solution is valid.

Fig.~\ref{fig:Fig1} shows the micromagnetic simulation of the above system. We see that the domain-wall velocity follows Eq.~\eqref{rdot} nearly perfectly and that velocity and $m_0$ approach the expected stationary values. It should be noted that our ansatz breaks down when $(\tilde{\gamma} G_2 a )^{-1}$
is not much smaller than $\lambda_w / \dot{r}_w$; however, a more detailed study of this regime is beyond the scope of this manuscript.

The typical values of the $\beta/G$-ratio in antiferromagnets is an interesting issue for future experiments. This ratio can be probed by measuring the domain-wall velocity as a function of the current density or by measuring the reciprocal process, which is voltage that is induced by a moving domain-wall. In the last case, one could initiate a domain-wall motion using a short current pulse and then measure the voltage echo from the reciprocal process. 

%%%%%%%%%%%%%%%%%%%%%%%%%%%
%\section{Conclusion}
%%%%%%%%%%%%%%%%%%%%%%%%%%%
In conclusion, we have derived a general phenomenological theory of current-induced dynamics
in antiferromagnets and have applied the theory to the study of current-induced domain-wall motion. We found that the domain-wall developed a net magnetic moment during the current-induced motion and that the domain-wall velocity was proportional to the ratio between the dissipative torque parameter and a damping parameter. 
%%%%%%%%%%%%%%%%%%%%%%%%%%%
%\section{acknowledgment}
%%%%%%%%%%%%%%%%%%%%%%%%%%%

This work was partially supported by the European Union FP7 Grant No.~251759 ``MACALO" (K. M. D. H. and A.B.), NSF Grant No.~DMR-0840965, and DARPA (Y.T.). We would like to thank Erik Wahlstr\"{o}m and Andr{\'e} Kapelrud for helpful discussions.


\begin{thebibliography}{99}
\bibitem{foot} It is useful to observe that calculating the forces $\mathbf{f}_n\equiv-\delta_{\mathbf{n}}F\to A\nabla^2\mathbf{n}$ and $\mathbf{f}_m\equiv-\delta_{\mathbf{m}}F\to-a\mathbf{m}+\mathbf{H}$ naively, i.e. not imposing the nonlinear constraints $|\mathbf{n}|=1$ and $\mathbf{m}\cdot\mathbf{n}=0$, results in the correct equations (\ref{LLG1}) and (\ref{LLG2}) [as well as our key equations (\ref{HTB1}) and (\ref{HTB2})], upon substitution. In the intermediate steps, such as Eqs.~(\ref{LL}), however, we will be using the fact that the proper forces are normal to the N{\'e}el order, $\mathbf{f}_n,\mathbf{f}_m\perp\mathbf{n}$.
\bibitem{foot2} Note that the second order differential equation for the N{\'e}el order dynamics depends on $\mathbf{m}(t=0)$ through the initial condition for $\mathbf{\dot{n}}(t=0)$.
\end{thebibliography}
\end{document}